\documentclass[12pt,preprint]{aastex}
\def\pmb#1{\setbox0=\hbox{$#1$}\kern-0.015em\copy0\kern-\wd0%
\kern0.03em\copy0\kern-\wd0\kern-0.015em\raise0.03em\box0}
\def\arcsec{\hbox{$^{\prime\prime}$}}

\shorttitle{Variations of scattering polarization in the line wings}
\shortauthors{Sampoorna et al.}
\begin{document}
\title{Origin of spatial variations of scattering polarization in the wings of the Ca\,{\sc i} 4227\,\AA\ line}
\author{M. Sampoorna,$^1$ J. O. Stenflo,$^2$ K. N. Nagendra,$^1$ M. Bianda,$^{2,3}$ 
R. Ramelli,$^3$ and L.~S. Anusha$^1$}
\affil{$^1$Indian Institute of Astrophysics, Koramangala,
Bangalore 560 034, India}
\affil{$^2$Institute of Astronomy, ETH Z\"urich, CH-8093 Z\"urich,
Switzerland}
\affil{$^3$Istituto Ricerche Solari Locarno, Via Patocchi, 6605 Locarno-Monti, Switzerland}

\affil{{\bf{Accepted in May 2009 for publication in The Astrophysical Journal}}}

\begin{abstract}
Polarization that is produced by coherent scattering can be modified by 
magnetic fields via the Hanle effect. This has opened a window to explorations 
of solar magnetism in parameter domains not accessible to the Zeeman effect. 
According to standard theory the Hanle effect should only be operating in 
the Doppler core of spectral lines but not in the wings. In contrast, our 
observations of the scattering polarization in the Ca\,{\sc i} 4227\,\AA\ 
line reveals the existence of spatial variations of the scattering 
polarization throughout the far line wings. This raises the question whether 
the observed spatial variations in wing polarization 
have a magnetic or non-magnetic origin. A magnetic origin may 
be possible if elastic collisions are able to cause sufficient frequency redistribution to make the Hanle effect effective in the wings without causing excessive collisional depolarization, as suggested by recent theories for partial frequency redistribution (PRD) with coherent scattering in magnetic fields. 

To model the wing polarization we bypass the problem of solving the full polarized radiative-transfer equations and instead apply an extended version of the technique based on the ``last scattering approximation'' (LSA). It assumes that the polarization of the emergent radiation is determined by the anisotropy of the incident radiation field at the last scattering event. We determine this anisotropy from the observed limb darkening as a function of wavelength throughout the spectral line. The empirical anisotropy profile is used together with the single-scattering redistribution matrix, which contains all the PRD, collisional, and magnetic-field effects. The model further contains a continuum opacity parameter, which increasingly dilutes the polarized line photons as we move away from the line center, and a continuum polarization parameter that represents the observed polarization level far from the line. This model is highly successful in reproducing the observed Stokes $Q/I$ polarization (linear polarization parallel to the nearest solar limb), including the location of the wing polarization maxima and the minima around the Doppler core, but it fails to reproduce the observed spatial variations of the wing polarization in terms of magnetic field effects with frequency redistribution. This null result points in the direction of a non-magnetic origin in terms of local inhomogeneities (varying collisional depolarization, radiation-field anisotropies, and deviations from a plane-parallel atmospheric stratification). 
\end{abstract}

 
\section{Introduction}
Coherent scattering on the Sun produces a linearly polarized spectrum that 
is as rich in spectral structures as the ordinary intensity spectrum, 
but has an entirely different appearance since the underlying physical 
processes are different \citep[cf.][]{ste04a}. This linearly polarized spectrum, 
which goes under the name the ``Second Solar Spectrum'', is modified by 
magnetic fields via the Hanle effect. It allows aspects of solar magnetism 
to be explored, which are not accessible by the Zeeman effect, in particular 
the vast amounts of ``hidden'' magnetic fields that have been revealed 
by Hanle-effect observations \citep[][see also Stenflo 2004b]{jtbetal04}. 

The atlas of the Second Solar Spectrum \citep{gan00,gan02,gan05} provides 
an overview of the linear polarization in lines observed 
near the solar limb, from the UV at 3160\,\AA\ to the red at 6995\,\AA. The 
largest degree of linear polarization in the visible spectrum is exhibited by the Ca\,{\sc i} 4227\,\AA\ line. Spatial variations of the linear 
polarization in the line core due to the Hanle effect have been observed in regions with
variable magnetic fields. Specropolarimetric measurements in this line
can be used to explore the magnetic field in the mid chromosphere 
\citep{bia98a,bia98b}. 

Recent observations by \citet{bia03} in the Ca\,{\sc i} 4227\,\AA\ line have revealed enigmatic behavior of the line wing polarization. These observations were made in active regions with the spectrograph slit perpendicular to the solar limb. They showed for the first time spatial variations of the linear polarization ($Q/I$ and $U/I$) in the far wings of the line, in contradiction with theoretical expectations. We will refer to this unexpected phenomenon as the ``$(Q/I, U/I)$ wing signatures''. This observation contradicts the long held standard theory for the Hanle effect, according to which the Hanle effect should only be effective in the line core but not in the line wings. The polarization is expected to approach the non-magnetic Rayleigh scattering limit in the line wings \citep[see][for details]{omo73,ste94,lan04}. In the present paper we report further such observations, this time done in quiet regions with the slit placed parallel to the nearest solar limb. Again we find similar spatial variations in the line wings as seen before in active regions. 

\citet{bia03} suggested a qualitative explanation for the observed $(Q/I,U/I)$ wing signatures in terms of partial frequency redistribution (PRD) and radiative transfer (multiple scattering) effects in the wings of strong resonance lines. Thus, in a balanced mixture of coherent and non-coherent scattering, it is possible to generate Hanle depolarization in the wings as follows\,: Hanle precession of the oscillating dipole moment is first generated near the resonance (in the line core), but gets shifted to a wing frequency by an elastic collision without destroying the atomic polarization. The atom subsequently emits the photon at the shifted frequency in the line wing. This process would be the source of the $(Q/I,U/I)$ wing signatures. Multiple scattering in the medium (due to finite monochromatic optical depth in the wings of strong resonance lines) may enhance this effect. \citet{nagetal02,nagetal03} showed through radiative transfer 
calculations that angle-dependent (AD) PRD is more efficient in generating 
shallow $(Q/I,U/I)$ wing peaks by this mechanism than the 
angle-averaged PRD. 

In the present paper we explore the above suggestions, using the last scattering 
approximation (LSA) instead of full radiative-transfer modeling, which is 
sufficient for our purpose of verifying the validity of the Hanle wing effect 
as an explanation of the observed wing polarization variations. This 
leads us to the rather unexpected conclusion that the Hanle effect cannot 
explain the observed wing effects, which suggests that the spatial variations of the wing 
polarization have a non-magnetic origin. For the scattering theory we use the 
recently developed Hanle-Zeeman angle-dependent PRD matrices for arbitrary magnetic fields 
\citep[][see also Sampoorna et al. 2009]{sametal07a,sametal07b}.

In \S~\ref{sec_obs} we present the observations of the $(Q/I,U/I)$ wing 
signatures. \S~\ref{sec_model} describes our theoretical model. The model fitting 
to the observed data is discussed in \S~\ref{sec_results}, while the conclusions 
are presented in \S~\ref{sec_conclu}.

\section{Observations of the $(Q/I,U/I)$ wing signatures}
\label{sec_obs} 

\subsection{Data acquisition}
Spectropolarimetric recordings of the full Stokes vector were obtained for the Ca\,{\sc i} 4227\,\AA\ line with the 45\,cm aperture Gregory Coud\'e Telescope (GCT) at IRSOL (Locarno, Switzerland). The ZIMPOL-2 polarimeter system was used
 \citep{ganetal04}, allowing highly precise measurements that are free from seeing-induced spurious effects, with an accuracy only limited by photon statistics. 
The observations were performed during the years 2005--2007 over 27 days. In total 
86 positions at different limb distances and at various latitudes on the solar disk were recorded, with the spectrograph slit parallel to the limb (which defines the positive Stokes $Q$ direction). Dark frames as well as flat fields were recorded before and/or after the observations. The polarimetric calibration and data reduction procedure has been described in \citet{ganetal04}.

The instrumental polarization in the GCT is mainly a function of declination and can be considered constant during a full observing day. Cross talk from Stokes $I$ to the other Stokes parameters is determined from flat field measurements in quiet regions at disk center. Since in the present analysis we are interested in the linear polarization away from active regions, we selected only regions where Zeeman-like signatures in the Stokes $V/I$ images are sufficiently small, so that the circular-to-linear polarization cross talk is negligible. We note that the circular-to-linear cross talk reaches its maximum at the solstices and is always smaller than 25\,\%\ \citep{rametal05}. 

It was carefully checked that the observed small signatures in the linear polarization were not of instrumental origin. To make sure that the observed signatures in the line wings do not originate from differential efficiencies of the different pixel rows of the ZIMPOL CCD, we alternated measurements by shifting the telescope image back and forth by 10\arcsec\ along the spectrograph slit direction. This could be achieved with the help of the automatic guiding system \citep{kuvetal03}.

\subsection{Observational results}
\begin{figure*}
\plotone{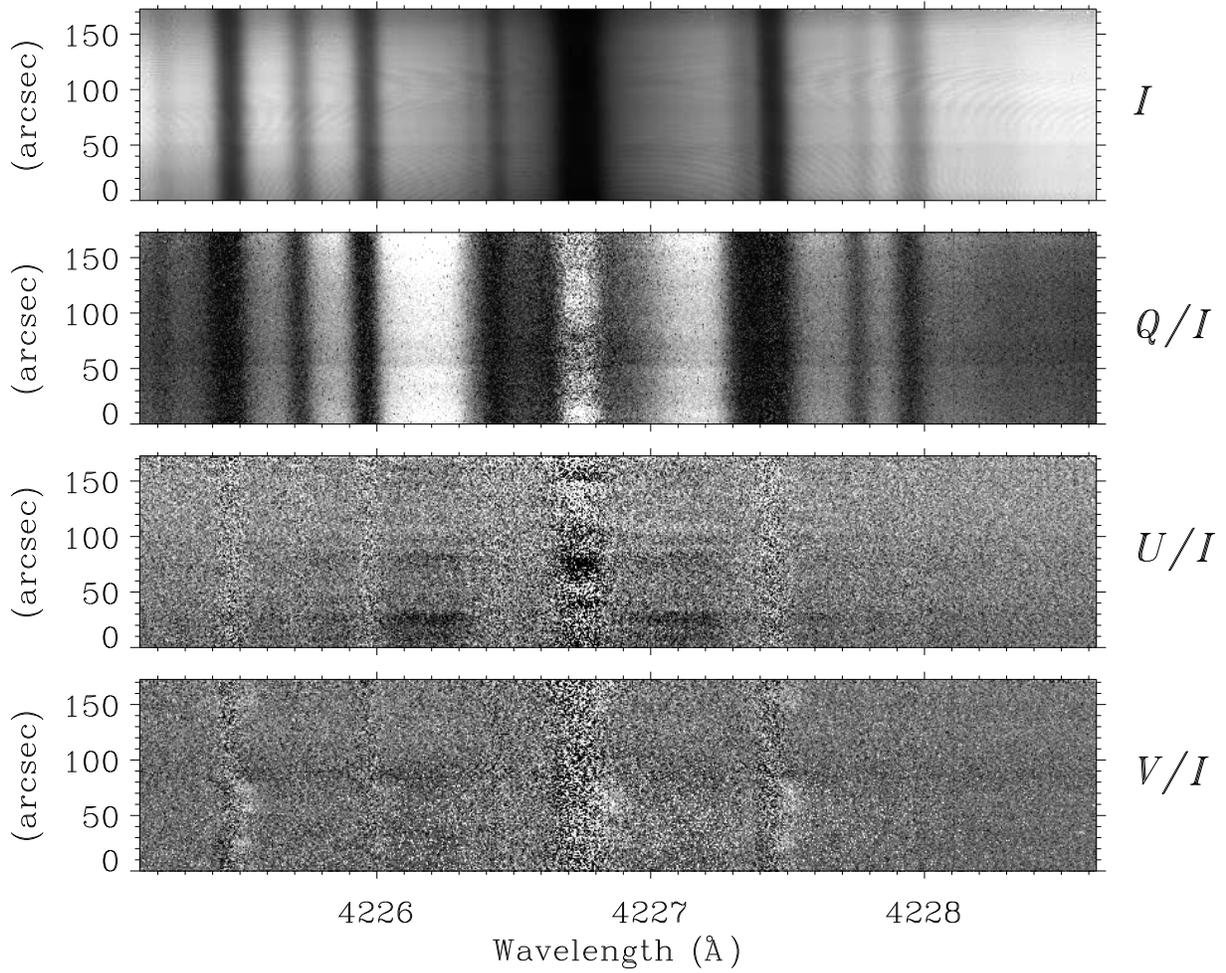}
\caption{CCD images of the Stokes parameters recorded with the 
spectrograph slit parallel to the limb (5\arcsec\  
inside the limb). Note the depolarization signatures in the 
$Q/I$ wings in the 55\arcsec\ - 75\arcsec\ spatial interval.  
Corresponding $U/I$ wing signatures are seen at the same wavelength 
positions, but at a different spatial location, for example 
in the 20\arcsec\ - 30\arcsec\ interval. 
}
\label{ccd_image}
\end{figure*}

Spatially varying linear polarization structures in the wings (of Stokes $Q$ and/or 
$U$) are found in 46 observations out of 86, thus in approximately half of all our 
recordings. This frequency of occurrence represents a {\it lower} limit, since non-optimum 
seeing conditions may smear the features to make them disappear below the noise level. Our 
observations thus show that such wing signatures (in $Q/I$ and/or $U/I$) are a very common 
phenomenon that is likely to be seen in the great majority of cases if the recordings are 
made with high spatial resolution. We illustrate in Fig.~\ref{ccd_image} a representative 
example of an observation obtained on 5 October 2007 near the heliographic north pole at 
about 5\arcsec\ inside the limb. The spectrograph slit width was 125 microns, which 
corresponds to 1\arcsec , and its length corresponded to about 170\arcsec. The total 
exposure time was 225\,s. A second measurement (not shown) taken with the slit shifted 
by 10\arcsec\ along the limb (and the slit) direction confirmed that the signatures 
were related to the solar positions and not to the position on the CCD.  

The intensity image shows the broad Ca\,{\sc i} line at 4226.74\,\AA\ with
blend lines in the wings. Due to the limb curvature, the largest limb
distance from the slit position is reached at around 85\arcsec\ from
the image bottom, while at 0\arcsec\ and 
170\arcsec\ the distance from the limb is minimum. This explains the brightness variation in the intensity 
image along the spatial direction.

The strong polarization signatures seen in $Q/I$ are 
due to scattering polarization. They occur both in the line core and wings, 
but decrease in amplitude in the very far wings. At the locations of the blend lines in the intensity image we also see the depolarization effects caused by 
these lines in the $Q/I$ image \citep{flusten01}. In the line core we see spatial variations (along the slit) both in $Q/I$ and $U/I$, which are caused by the Hanle effect \citep{bia98a,bia98b,bia99} in the presence of magnetic fields in the mid chromosphere. The $Q/I$ polarization in the Ca\,{\sc i} 4227 line wings 
increases at the edges of the spectrograph slit, since the edges are located closer to the solar limb. Narrow horizontal strips that can be seen for instance in the interval 55\arcsec\ to 75\arcsec\ from the bottom of the $Q/I$ image show enigmatic `depolarization 
signatures' in the line wings, which we want to explore. 

The $U/I$ image shows features in the line core that are due to 
Hanle rotation of the polarization plane. However, structures can also be observed 
in the line wings but at different spatial locations than the line core 
features, i.e., in the intervals 20\arcsec - 30\arcsec, 40\arcsec - 
50\arcsec, and 70\arcsec -80\arcsec. We note that there are hardly 
any spatial correlations between the $Q/I$ and $U/I$ wing signatures.

\begin{figure*}
\plotone{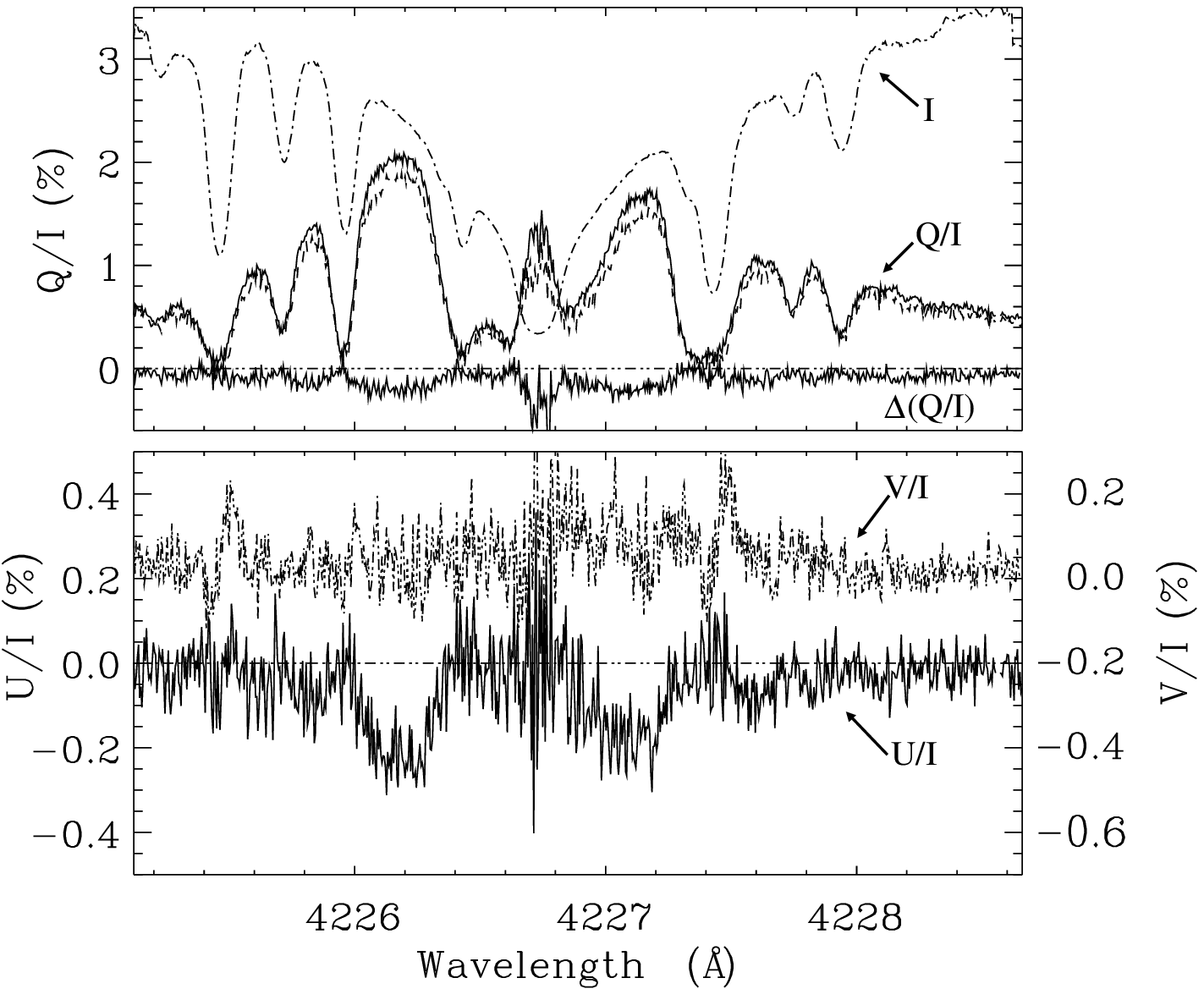}
\caption{Stokes $I$, $Q/I$, $U/I$, $V/I$ profiles extracted from 
Fig.~\ref{ccd_image}. The heliocentric angle corresponds to $\mu=0.1$. Note the depolarization  in the $Q/I$ wings and the $U/I$ signatures at the corresponding wavelength positions. In the present paper we refer to them together as the ``$(Q/I,U/I)$ wing signatures''. 
}
\label{obs_res2}
\end{figure*}
 
Figure~\ref{obs_res2} shows profiles averaged over different spatial 
intervals. In the top panel the dot-dashed line represents the  intensity profile, in arbitrary units, averaged over a 40\arcsec\ broad interval around the middle of the spectrograph slit. The $Q/I$ profile shown by the solid line is obtained by averaging outside the interval where depolarization in the line wings is observed, while the dashed line represents the profile obtained by averaging in the interval between 55\arcsec\ - 75\arcsec. The difference between the two averaged $Q/I$ profiles is shown by the $\Delta (Q/I)$ line.

In the bottom panel of Fig.~\ref{obs_res2} the solid line represents the 
$U/I$ profile averaged over the spatial interval 29\arcsec - 
40\arcsec. Its shape resembles a negative version of the $Q/I$ profile. 
Note that the Stokes $U/I$ wing signals of Figs.~\ref{ccd_image} and \ref{obs_res2}
cannot be due to the possibility that the positive Stokes $Q$ direction may not be
exactly parallel to the nearest limb, because in that case the wing polarization in
$U/I$ would simply be proportional to the wing polarization in $Q/I$ all along the slit,
which is not the case. It is important to note that in our observations the $Q/I$ and
$U/I$ wing polarizations never vary in synchrony along the slit but have different
spatial structures.

The $V/I$ profile represented by the dotted line is averaged in the 
interval where the largest polarization amplitudes are seen (24\arcsec - 
45\arcsec). Only very faint signatures can be seen in the Fe\,{\sc i} lines 
at 4225.46 and 4227.44\,\AA. The absence of large signals in $V/I$ 
shows that we are observing a solar region with very weak longitudinal 
magnetic field components.

\section{Theoretical model}
\label{sec_model}
Quantitative modeling of the scattering polarization with the Hanle effect 
requires the solution of the relevant radiative transfer problem. Such 
calculations have been done by \citet[][and references cited therein]
{fau92,fauetal09} and \citet{reneetal05,reneetal06}, who use angle-averaged   
(AA) PRD. Extensive radiative transfer 
calculations have also been done by Trujillo Bueno and 
co-workers \citep[see][and references cited therein]{jtb09} to 
model the scattering line polarization and the Hanle effect in terms of the 
complete frequency redistribution (CRD) approximation, but taking into account 
atomic level polarization in multilevel atomic systems. 

For exploratory purposes we can avoid such full-scale radiative transfer modeling by using 
semi-empirical approaches 
in terms of the last scattering approximation (LSA), which has proven successful in the past \citep{ste82}. In the present section we describe how LSA in combination with the full redistribution matrix for arbitrary magnetic fields can be used 
to model the Second Solar Spectrum. We 
illustrate this approach by applying it to the Ca\,{\sc i} 4227\,\AA\ line 
observations presented in \S~\ref{sec_obs}. 

One of the earliest works on modeling the linearly polarized solar spectrum with LSA dates back to \citet{ste80}, where the observed Ca\,{\sc ii} H and K line 
wing polarization that exhibited a quantum-interference signature extending over about 200\,\AA\ was fitted. In a later paper \citet{ste82} extended the LSA concept 
to interpret Hanle polarization observations of several spectral lines in terms of a 
micro-turbulent magnetic field, which allowed the strength of the ``hidden'' tangled fields to be estimated for the first time. 

In the present paper we extend this approach to the exploration of the 
Hanle effect for partially resolved magnetic fields, which produce 
signatures in both $Q/I$ and $U/I$. In the following subsections we revisit 
each of the important ingredients of Stenflo's method, for the purpose 
of clarity and completeness. 

\subsection{Last scattering approximation (LSA)} 
The concept of LSA is particularly useful in 
astronomical contexts where either (i) the geometry is too complicated or 
(ii) the relevant transfer equation is difficult to solve 
computationally. Recent applications of this concept in the modeling of scattering polarization can be found in 
\citet[][modeling of molecular emission lines]{fauarn02}; 
\citet[][modeling the solar continuum polarization]{ste05}; 
\citet[][modeling of the Ba {\sc ii} D$_2$ line]{belletal07}; and 
\citet[][Hanle scattering in random magnetic fields - where a variant of the LSA idea is presented]{hfetal09}. 

LSA exploits the fact that the polarization of the radiation that 
escapes the atmosphere is mainly determined by the anisotropy of the 
radiation field at the place where the last scattering process takes place. Since the polarization 
amplitudes in the lines are small, the polarization of the incident radiation 
at the last scattering event can be neglected. In other words, the emergent 
polarization is produced in a single scattering event (the very last one) 
rather than through multiple scattering within the atmosphere. 

In the practical application of this idea the most important limiting assumption 
is that we base the radiation-field anisotropy that we apply to the 
single-scattering event, on the observed limb darkening function which represents 
the top of the atmosphere. 
In reality most of the observed photons originate from $\tau\approx\mu$. This 
difference however becomes negligible for extreme limb observations 
($\mu \to 0$), for which the emergent radiation represents the topmost 
layers of the atmosphere ($\tau \approx 0$). If frequency coherent scattering in the laboratory frame is assumed, LSA allows us to write the 
emergent $Q/I$ polarization for non-magnetic scattering in a neatly factorized 
form: 
\begin{equation}
\label{qbyi_coh}
{Q\over I} \equiv P = W_{\rm 2,eff}\, k_{\rm G,\lambda}(\mu)\, k_c,
\end{equation}
where $\mu=\cos\theta$, with $\theta$ being the heliocentric angle 
\citep[see][]{ste94}. Here $W_{\rm 2,eff}$ is the effective atomic 
polarizability factor, which is unity for the Ca\,{\sc i} 4227\,\AA\ line, 
since it behaves like classical dipole scattering. The blend lines on 
the other hand generally do not polarize, which means that they have a 
$W_2=0$. Their non-polarizing opacity therefore dilutes the polarized 
Ca\,{\sc i} 4227 line photons, making the ``effective polarizability'' 
$W_{\rm 2,eff}$ much smaller than unity inside the blend lines. 
$k_{\rm G,\lambda}(\mu)$ is a geometric depolarization factor that depends 
on the anisotropy of the radiation field, and the line of sight with 
respect to the local normal. It describes the depolarization 
caused by the angular integration over the incident radiation. In practice this factor is determined from the observed center-to-limb variation (CLV) 
of Stokes $I$ (see \S~\ref{kg_clv}). $k_c$ is the collisional 
depolarization factor given by 
$\Gamma_{\rm R}/(\Gamma_{\rm R}+\Gamma_{\rm E})$, where $\Gamma_{\rm R}$ 
and $\Gamma_{\rm E}$ are radiative and elastic collision rates. 

\subsection{Empirical determination of the anisotropy factor $k_{\rm G,\lambda}(\mu)$}\label{kg_clv}
$k_{\rm G,\lambda}(\mu)$ can be determined using the observed limb darkening 
function, which is defined as follows\,:
\begin{equation}
\label{limb_dark_func}
c_{\lambda}(\mu) \equiv {I_{\lambda}(\mu) \over I_\lambda (\mu=1)}.
\end{equation}
However, what is actually observed is a series of unnormalized spectra 
$b(\mu) I_{\lambda}(\mu)$, where $b(\mu)$ is some arbitrary scaling factor 
that is different for each $\mu$. To eliminate the arbitrary $b(\mu)$ 
we normalize each spectrum to the continuum intensity. Since a true 
continuum is usually not recorded, we choose a reference wavelength $\lambda_{\rm ref}$ 
at which we are as close to the continuum as we can be. Further we assume that 
the CLV at that reference wavelength is the same as the CLV of the continuum, 
namely
\begin{equation}
\label{ref-wave_cont}
{I_{\lambda_{\rm ref}}(\mu) \over I_{\lambda_{\rm ref}} (\mu=1)} \approx 
{I_c(\mu) \over I_c(\mu=1)}.
\end{equation}
The observed quantity that we have to work with is 
\begin{equation}
\label{iobs}
I_{\rm obs,\lambda} (\mu) = {b(\mu)\, I_{\lambda}(\mu) \over b(\mu)\, 
I_{\lambda_{\rm ref}}(\mu)},
\end{equation}
for each spectrum, so that $b(\mu)$ divides out. We can then write the limb 
darkening function as 
\begin{equation}
\label{limb_dark_func2}
c_{\lambda}(\mu) = {I_{\rm obs,\lambda}(\mu) \over I_{\rm obs,\lambda} (\mu=1)}
\ \ {I_c(\mu) \over I_c(\mu=1)}.
\end{equation}
For the limb darkening function of the continuum around the 4227\,\AA\ line we use the following 
analytical representation
\begin{equation}
\label{cont_clv}
{I_c(\mu) \over I_c(\mu=1)} = 1 - a_{0,c} - a_{1,c} +a_{0,c}\, \mu + a_{1,c}\, \mu^2,
\end{equation}
where $a_{0,c}$ and $a_{1,c}$ are fit parameters taken from \citet{pierce00}. 
The limb darkening function $c_\lambda(\mu)$ determined from the observed data is 
then fitted by the following function
\begin{equation}
\label{wave_clv}
f_\lambda(\mu) = 1 - a_{0,\lambda} - a_{1,\lambda} +a_{0,\lambda}\, \mu + a_{1,\lambda}\, 
\mu^2.
\end{equation}
Least squares fitting of $c_\lambda(\mu)$ in terms of $f_\lambda(\mu)$ gives us the values of the coefficients $a_{0,\lambda}$ and $a_{1,\lambda}$. 
It is important to note that the $\lambda_{\rm ref}$ chosen should 
be kept the same for all the recordings of Stokes $I$ with different 
$\mu$ values. 

For use with LSA we obtain the geometric depolarization factor $k_{\rm G,\lambda}(\mu)$ by multiplying the Rayleigh phase 
matrix with an unpolarized Stokes vector $(I,\ 0,\ 0,\ 0)^{\rm T}$ and 
integrating over all the incoming angles. This gives \citep[see][]{ste82,ste05} 
\begin{equation}
k_{\rm G,\lambda}(\mu) = G_{\lambda} (1-\mu^2) / I_{\lambda}(\mu),
\label{kg_exp}
\end{equation}
where
\begin{equation}
G_{\lambda} = {3\over 16} \int_{-1}^{+1} (3\mu^{\prime^2}-1) 
I_{\lambda}(\mu^\prime) {\rm d}\mu^\prime.
\label{glambda}
\end{equation}
Note that our definition of $G_{\lambda}$ differs from that of \citet{ste82,ste05} 
only by a negative sign. This sign change is made to account for the circumstance that 
the positive $Q$ direction in the theoretical
calculations of scattering matrices is defined to be perpendicular to the limb, while it
is defined to be parallel to the limb in the observations.
In \citet{ste05} this sign change has been made in the final expression for $G_{\lambda}$ 
presented in that paper (see his Eq.~(31)). For convenience of our purposes, it is 
sufficiently accurate to assume that 
the actual limb darkening $(I_{\lambda}(\mu)/I_{\lambda}(\mu=1))$ can be 
represented by a parabolic 
type function $f_{\lambda}(\mu)$ as given by Eq.~(\ref{wave_clv}). This allows us to perform the integration in Eq.~(\ref{glambda}) 
analytically. $f_{\lambda}(\mu)$ is defined for the outwards hemisphere 
(positive $\mu$) only. For the inwards hemisphere (negative $\mu$) it is 
assumed to be zero, which is a valid assumption at the surface 
($\tau_\lambda=0$). Thus we obtain 
\begin{eqnarray}
&&k_{\rm G,\lambda}(\mu) = \left({3a_{0,\lambda} \over 64} + {a_{1,\lambda}
\over 20}\right)\nonumber \\ &&\times{(1-\mu^2) \over 
1 - a_{0,\lambda} - a_{1,\lambda} +a_{0,\lambda}\, \mu + a_{1,\lambda}\, \mu^2}. 
\label{kg_exp2}
\end{eqnarray}

Notice that the determination of $a_{0,\lambda}$ and $a_{1,\lambda}$ and thus of $k_{\rm G,\lambda}(\mu)$ crucially depends on the 
observed CLV of Stokes $I$. Hence we made a dedicated set of 
observations on January 10, 2009, to record the intensity spectrum with great 
precision at several $\mu$ positions. The slit was oriented parallel to the 
geographic north pole and placed at different $\mu$ positions. Since the intensity flat field of the detector in the spectrograph focus is very important for this type of observation, we applied a careful flat-fielding procedure by combining images taken while the telescope was moving in a random pattern around disk center and the spectrograph grating position was unchanged, with images taken while the grating rotated to smear the spectrum. The $\mu$ positions were calculated to a high degree of accuracy using 
three methods, namely (i) the slit position calculated from the digitized slit 
jaw image, (ii) the slit position calculated from the guiding system 
\citep[primary image guider;][]{kuvetal98}, and (iii) the slit position 
calculated with the aid of the 
encoder system, giving the telescope position in right ascension and declination. The 
various $I_{\lambda}(\mu)$ recordings were brought to a common wavelength scale through 
interpolation, so that the wavelength position of all the blend lines 
perfectly match for the different $I_{\lambda}(\mu)$ observations. Otherwise, due to 
small wavelength drifts between the different $I_{\lambda}(\mu)$ recordings one 
can get spurious peaks in $k_{\rm G,\lambda}(\mu)$ due to gradient effects in the vicinity of the blend lines. 

\begin{figure}[ht]
\plotone{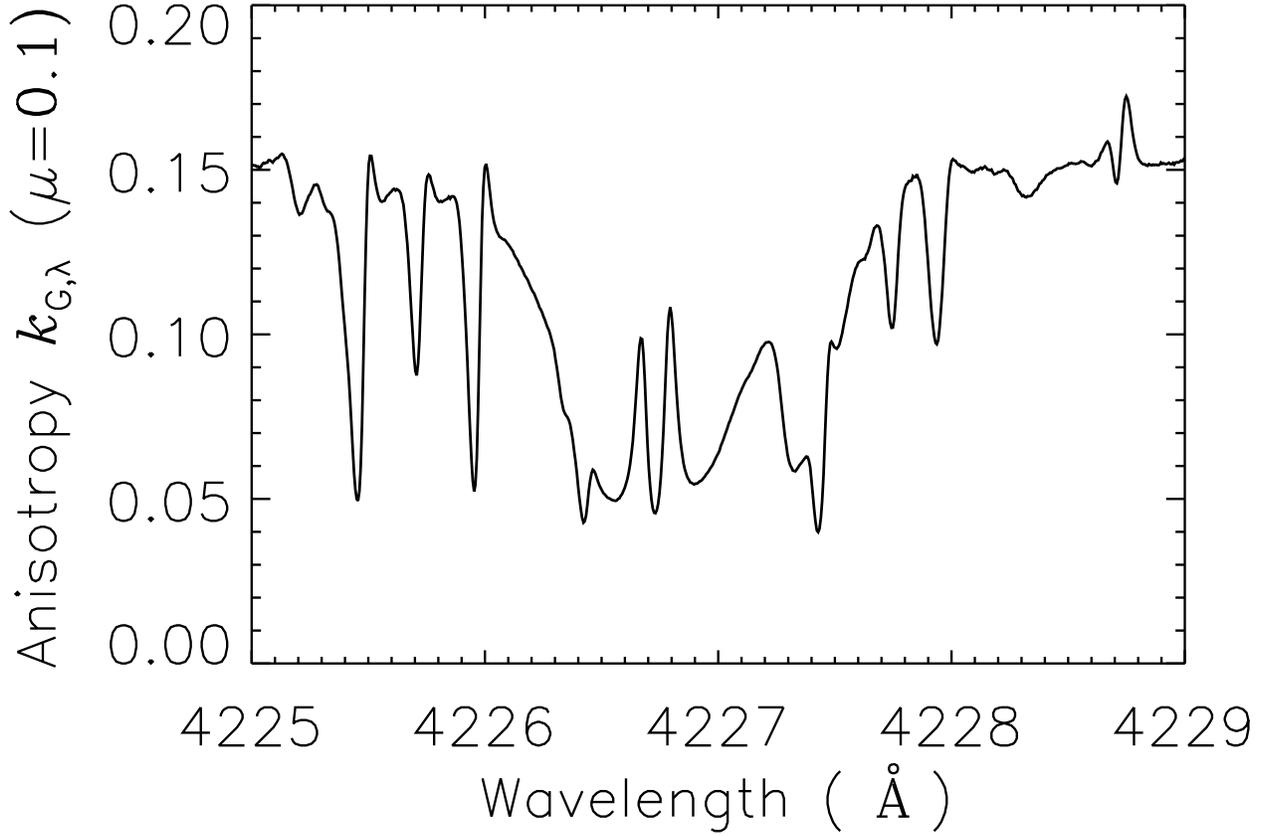}
\caption{Plot of the geometric depolarization or anisotropy factor 
$k_{\rm G,\lambda}(\mu)$ as a function of $\lambda$ for disk position $\mu=0.1$.}
\label{kg_plot}
\end{figure}

Fig.~\ref{kg_plot} shows a plot of the geometric depolarization 
factor $k_{\rm G,\lambda}(\mu)$ for $\mu=0.1$. It can also be called 
`anisotropy factor' as it is largely governed by $G_\lambda$ (see Eq.~(\ref{glambda})). The resemblance of $k_{\rm G,\lambda}(\mu)$ to the 
$I_{\lambda}(\mu)$ spectra is very striking in the far wings. However, it also has a distinctive 
shape in the core and wings of the Ca\,{\sc i} 4227\,\AA\ line. 
The anisotropy factor has a minimum in both the core of the main line and the cores of the surrounding blend lines. It is largest in the line wings, where it reaches a nearly 
constant value. We note that our anisotropy plot resembles the anisotropy curve $J^2_0/J^0_0$ shown as the solid line in Fig.~5 (right bottom panel) of \citet{reneetal05} for the same line. 

\subsection{Atomic and atmospheric data}
Ca\,{\sc i} 4227\,\AA\ is a resonance line for which the lower level is the 
ground state, and the coupling to other bound states of Ca\,{\sc i} may be 
neglected \citep[see][]{fau92}. Therefore a 2-level model atom is a reasonably good approximation. The ($\, ^1\!S_0 \to\,^1\!P_1 \to\, ^1\!S_0$) scattering transition produces a triplet line in the presence of strong magnetic fields. In weak fields the partially split $m$-states coherently superpose (interfere) to give rise to the Hanle effect. We take the required atomic data for 
this line from \citet{fau92}. The radiative width is $\Gamma_{\rm R} = 2.18\times 
10^8\ {\rm s}^{-1}$. The Doppler width is given by 
\begin{equation}
\label{dop_width}
\Delta{\lambda_{\rm D}} = {\lambda_0 \over c} \sqrt{{2 k T \over M_a} +
 v^2_{\rm turb}} \ ,
\end{equation}
where $c$ is the speed of light, $k$ the Boltzmann constant, $M_a$ the 
mass of a Ca\,{\sc i} atom. For a temperature $T=6000$\,K\ and a turbulent 
velocity $v_{\rm turb}=2$\,kms$^{-1}$, the Doppler width is 35.9\,m\AA. 
The corresponding damping parameter $a_{\rm R} = \Gamma_{\rm R} / (4\pi
\Delta\nu_{\rm D}) = 2.8\times 10^{-3}$. 

\subsection{Model for the non-magnetic scattering polarization}
\label{sec_nonmag_model}
It is well known that strong resonance lines like Ca\,{\sc i} 4227\,\AA\ can 
be modeled only when PRD effects are taken into account. Therefore we need to use appropriate PRD matrices in our LSA approach. The relevant expressions 
can either be taken from  \citet{domhub88}, or computed from our Hanle-Zeeman 
theory with $B=0$ \citep[see][]{sametal07a,sametal07b}. Further we now need to 
generalize Eq.~(\ref{qbyi_coh}), which was formulated for frequency coherent 
scattering in the laboratory frame by \citet{ste82}. 
The expression for the line contribution to the 
linear polarization $Q/I$ is given by 
\begin{equation}
\label{pqline_prd}
P_{Q,{\rm line}} = {\int R_{21}(\lambda,\lambda^\prime,\Theta)\, 
k_{\rm G,\lambda^\prime}(\mu)\, I_{\lambda^\prime}(\mu=1)\,{\rm d}\lambda^\prime 
\over \int R_{11}(\lambda,\lambda^\prime,\Theta)\, 
I_{\lambda^\prime}(\mu=1)\,{\rm d}\lambda^\prime}.
\end{equation}
Here $R_{i1}(\lambda,\lambda^\prime,\Theta)$ are 
the redistribution matrix elements for $i=1,2$. They depend on the scattering angle 
\begin{equation}
\Theta = \cos^{-1}\left[\cos\theta\cos\theta^\prime + 
\sin\theta\sin\theta^\prime \cos(\phi-\phi^\prime)\right],
\label{scat_angle}
\end{equation}
where $(\theta, \phi)$ and $(\theta^\prime, \phi^\prime)$ are respectively the 
outgoing and incoming ray directions with respect to the atmospheric normal. 

The expression for $P_{Q,{\rm line}}$ (i.e., Eq.~(\ref{pqline_prd})) can be 
justified as follows\,: According to LSA the emergent polarization is produced by the 
last scattering event. The incident radiation is unpolarized, so we only need 
to consider the single-scattering redistribution matrix elements 
$R_{i1}(\lambda,\lambda^\prime,\Theta)$ as done in Eq.~(\ref{pqline_prd}). The 
angular integration over incoming angles is avoided by applying the 
$k_{\rm G,\lambda^\prime}(\mu)$ factor, which embodies the effect
of the anisotropy of the incident radiation field. 
Thus the matrix element $R_{21}$ simply needs to be scaled with this factor 
and integrated over all incoming wavelengths. 
Conceptually we have here decomposed the angular integral into two parts. 
The first part consists of a unidirectional delta function scaled with 
$k_{\rm G,\lambda^\prime}(\mu)$. 
The second part is isotropic and vanishes on angular integration with $R_{21}$.  
Only the delta function contributes to the polarization. 

When CRD is assumed, we have $R_{i1}(\lambda,\lambda^\prime,\Theta)=
H(\Delta\lambda^\prime,a)H(\Delta\lambda,a)P_{i1}(\Theta)$ where $H(\Delta\lambda,a)$ 
is the Voigt function (see below) and $P_{i1}(\Theta)$ are the non-magnetic Rayleigh 
phase matrix elements \citep[see for e.g.,][for their expressions]{ste94}. Thus under CRD 
Eq.~(\ref{pqline_prd}) becomes wavelength independent and therefore is valid only in the 
line core (see \S~\ref{sec_prd}). 

For the calculations presented in this paper, we choose $\cos\theta=\mu=0.1$, for 
which the observations were made. For the scattering geometry we use $\cos\theta^\prime=1$ 
and $\phi=\phi^\prime=0$. The collisional depolarization factor $k_c$ is self-consistently 
contained in $R_{21}$ and $R_{11}$ through proper branching ratios. 
We use the angle-dependent PRD matrices 
for all our modeling purposes, unless stated otherwise. 

In order to model the observations we also need to take into account the contributions of the continuum, namely the continuum opacity and the continuum polarization. They are 
included in our model as follows\,:
\begin{equation}
\label{qbyi_nonmag_model}
{Q\over I} = S \left[ P_{Q,{\rm line}} {H(\Delta\lambda,a)\over H(\Delta\lambda,a) + C} 
 + P_c {C \over H(\Delta\lambda,a) + C} \right].
\end{equation}
$H(\Delta\lambda,a)$ is the Voigt function that 
describes the absorption probability for the Ca\,{\sc i} 4227\,\AA\ line, with 
damping parameter $a$ given by 
\begin{equation}
\label{total_damp}
a = {\Gamma_{\rm R}+\Gamma_{\rm I}+\Gamma_{\rm E}\over 4\pi\Delta\nu_{\rm D}} = 
a_{\rm R} \left[1+{\Gamma_{\rm I}+\Gamma_{\rm E}\over \Gamma_{\rm R}}\right].
\end{equation}
Since the inelastic collision rate $\Gamma_{\rm I}\ll \Gamma_{\rm E}$, 
we set $\Gamma_{\rm I}=0$. For a given 
choice of $\Gamma_{\rm E}/\Gamma_{\rm R}$, the free parameters 
of our model are $S,\ C,\ P_c$. The global scaling parameter $S$ is adjusted such that 
the amplitude of the modeled $Q/I$ blue wing maximum agrees with the observed value. Ideally $S$ should be close to unity. Large departures from unity 
may be due to transfer effects and/or collisional depolarization. The 
continuum opacity parameter $C$ plays the dominant role. It allows us to 
reproduce the overall 
shape of the observed $Q/I$ in the near and far wings of the Ca\,{\sc i} 
4227 line. In particular it determines the wavelength positions where the 
maximum wing polarization is reached, beyond which the polarization starts 
to decline again. The continuum 
polarization $P_c$ only determines the asymptotic shape of $Q/I$ in the very far wings and does not 
play an important role here. $P_c$ is fixed by the asymptotic 
behavior of $Q/I$ far from the line center. 
From the atlas of \citet{gan02} we determine the ratio 
$r_{\rm obs} = (P_c)_{\rm obs}/P_{\rm wing, max(4227)}$. In our modeling we 
always choose $P_c$ such that $r_{\rm model} = r_{\rm obs}$. The fitting procedure 
is as follows\,:
\begin{itemize}
\item[(1)] Choose a given value of $C$.
\item[(2)] Choose a $P_c$ that makes $r_{\rm model} = r_{\rm obs}$.
\item[(3)] Find a value of $S$ that makes $(Q/I)_{\rm wing, max}$ of the model 
agree with the observed value.
\end{itemize}
Iterate (1) to (3) until the best fit is obtained. The fitting 
procedure is repeated for different choices of $\Gamma_{\rm E}/\Gamma_{\rm R}$, so that we obtain $S,\ C$, and $P_c$ as functions of $\Gamma_{\rm E}/\Gamma_{\rm R}$. 

\subsection{Model for magnetic scattering polarization}
\label{sec_hanle_model}
Assuming that the non-magnetic anisotropy factor $k_{\rm G,\lambda}(\mu)$ is still valid in the presence of weak fields, we extend the model of 
\S~\ref{sec_nonmag_model} to include the Hanle effect. We further assume 
that the same $k_{\rm G,\lambda}(\mu)$ can be used for both $Q/I$ and $U/I$. 
This implies that the decomposition of the angular integral into a 
contribution from a unidirectional delta function scaled with 
$k_{\rm G,\lambda}(\mu)$, while the rest represents isotropic scattering, 
is equally valid for both $Q/I$ and $U/I$. This is a reasonably  
good approximation, but we plan to test it in future work. Thus the model 
$U/I$ profile for the Hanle effect is given by
\begin{equation}
\label{ubyi_hanle_model}
{U\over I} = S P_{U,{\rm line}} \,{H(\Delta\lambda,a)\over H(\Delta\lambda,a) +  C}\,,
\end{equation}
where
\begin{equation}
\label{puline_prd}
P_{U,{\rm line}} = {\int R_{31}(\lambda,\lambda^\prime,\Theta)\, 
k_{\rm G,\lambda^\prime}(\mu)\, I_{\lambda^\prime}(\mu=1)\,{\rm d}\lambda^\prime \over \int R_{11}(\lambda,\lambda^\prime,\Theta)\, 
I_{\lambda^\prime}(\mu=1)\,{\rm d}\lambda^\prime}.
\end{equation}
$Q/I$ is given by Eqs.~(\ref{qbyi_nonmag_model}) and (\ref{pqline_prd}), but now 
$R_{21}$ and $R_{11}$ contain magnetic field contributions. For the scattering 
redistribution matrix elements $R_{i1}$ with $i=1,2,3$ we use the Hanle-Zeeman 
theory \cite[see][]{sametal07a,sametal07b}, although we may also use approximation-II 
of \citet{bom97}. Note that in a plane-parallel atmosphere the continuum 
polarization contribution to the $U/I$ is zero. However, as shown by \citet{js09}, in a 
real 3D model atmosphere this contribution to $U/I$ is actually a non-zero quantity whose local 
value fluctuates in sign at the spatial scales of the horizontal inhomogeneities 
that produce symmetry breaking in the radiation field. The fitting procedure for the weak 
magnetic field case is as follows\,:  
\begin{itemize}
\item[(1)] First the model parameters $S,\ C,\ P_c$ for different choices of 
$\Gamma_{\rm E}/\Gamma_{\rm R}$ are fixed by fitting the $Q/I$ observed 
in a quiet region (see \S~\ref{sec_nonmag_model}). 
\item[(2)] For these fixed parameters we use the magnetic 
redistribution matrix to compute $U/I$ and $Q/I$ for various choices of $\Gamma_{\rm E}/\Gamma_{\rm R}$ 
and the vector magnetic field ${\pmb{B}}$ parameters. 
\item[(3)]  We then explore which combinations of ${\pmb{B}}$ and 
$\Gamma_{\rm E}/\Gamma_{\rm R}$ best reproduce the observations. 
\end{itemize}

\section{Results and Discussion}
\label{sec_results}

\subsection{Model fit of the non-magnetic $Q/I$}
Using the procedure described in \S~\ref{sec_nonmag_model}, we fit the observed non-magnetic $Q/I$ data shown as the solid line in the top panel of 
Fig.~\ref{obs_res2}. We recall that such a fit fixes the values of 
the free parameters $S$, $C$ and $P_c$. 
\begin{figure}[ht]
\plotone{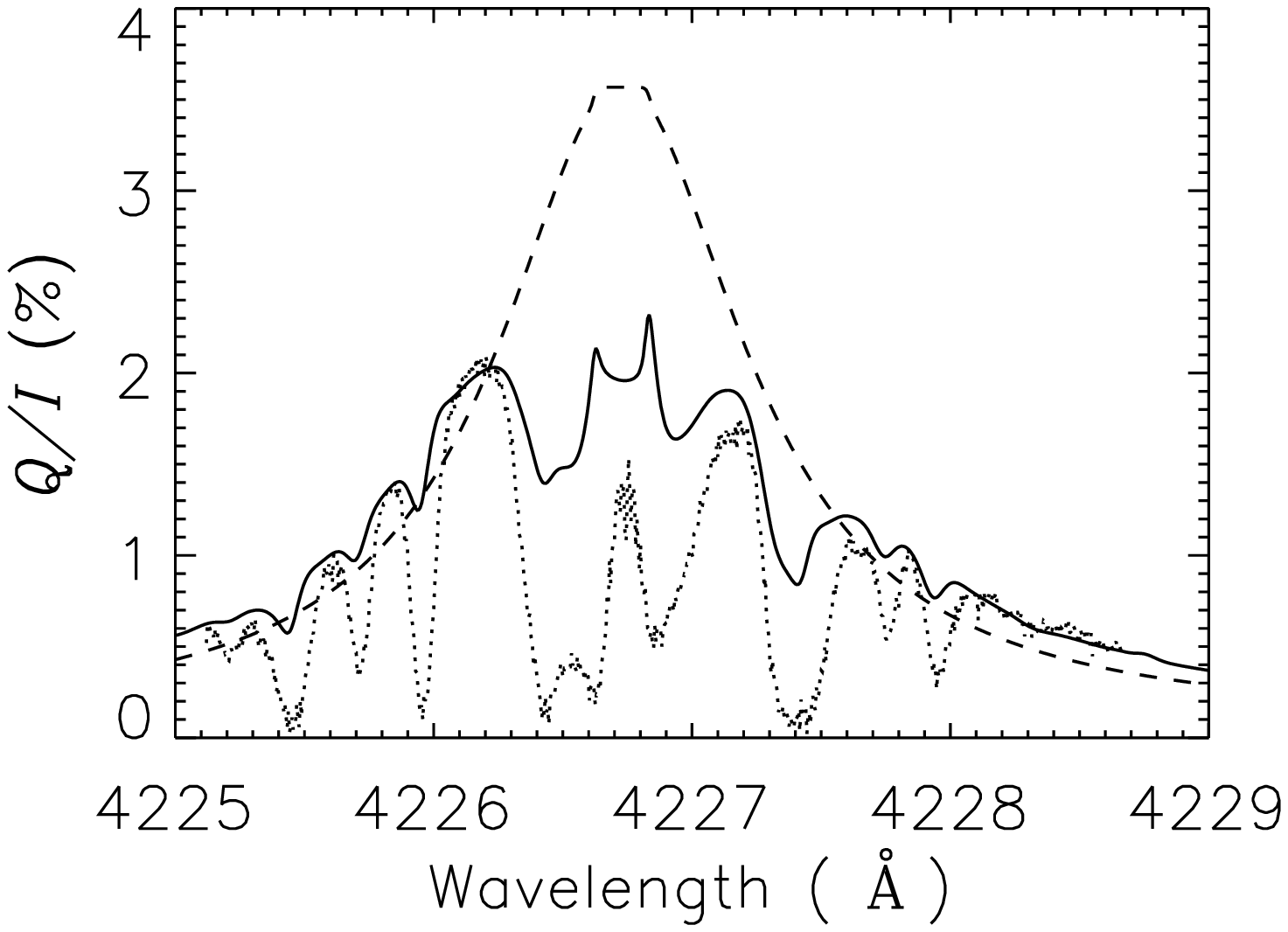}
\caption{Role of $k_{\rm G,\lambda}(\mu)$. The solid line is the 
model profile computed using pure $R_{\rm II-AD}$ type redistribution and 
$k_{\rm G,\lambda}(\mu)$ determined from observations. The dotted line is 
the observed $Q/I$ in the non-magnetic region (same as the solid line 
in the top panel of Fig.~\ref{obs_res2}). The dashed line is the model 
profile computed using pure $R_{\rm II-AD}$ type redistribution but with 
a flat $k_{\rm G,\lambda}(\mu) = 0.112$ for all wavelengths. The 
free parameters obtained from the model fit are 
$S=0.325$, $C=9.7\times 10^{-5}$ and $P_c=0.25\ \%$.
}
\label{flatkgvskg}
\end{figure}

\subsubsection{Role of the geometric depolarization factor $k_{\rm G,\lambda}(\mu)$}
To illustrate the important role of the anisotropy factor $k_{\rm G,\lambda}(\mu)$ we present in Fig.~\ref{flatkgvskg} the model profiles computed 
using $k_{\rm G,\lambda}(\mu)$ determined from observations (solid line) 
and computed with a flat $k_{\rm G,\lambda}(\mu)$ (dashed line). For 
the flat $k_{\rm G,\lambda}(\mu)$ we choose the value of non-flat 
$k_{\rm G,\lambda}(\mu)$ at the wavelength of the $Q/I$ blue wing peak, and keep it constant for all other wavelengths. To compute the model profiles 
in Fig.~\ref{flatkgvskg} (solid and dashed lines) we have used the collisionless PRD matrix (i.e., angle-dependent pure $R_{\rm II-AD}$ type redistribution). 
The free parameters obtained by the model fit are $S=0.325$, 
$C=9.7\times 10^{-5}$, and $P_c=0.25\ \%$. 

Clearly the entire structuring of the $Q/I$ model profile, with the 
minima around the Ca\,{\sc i} 4227\,\AA\ Doppler core and the blend 
line depressions are all related to the $k_{\rm G,\lambda}(\mu)$ structure 
(see Fig.~\ref{kg_plot}). The blend line minima of the model 
profile (solid line) are less deep than in the observed spectrum (dotted 
line). The main reason why the computed blend lines are not sufficiently
deep in $Q/I$ is not due to $k_{\rm G,\lambda}(\mu)$ alone, but because 
we have disregarded that the blend line opacities can have intrinsic 
polarizability $W_2=0$ and thus dilute the Ca\,{\sc i} 4227 line photons with 
unpolarized photons. We have chosen to ignore this property here, to avoid 
introducing more free parameters and keep the model as simple as possible. 

\subsubsection{The role of partial frequency redistribution}
\label{sec_prd}
The use of frequency coherent scattering in the laboratory frame 
(static atoms) is physically incorrect due to Doppler redistribution. It 
can still be used as a good approximation in the line wings, but 
becomes invalid in the line core. For a correct treatment we need 
partial frequency redistribution (PRD). 

\begin{figure}[ht]
\plotone{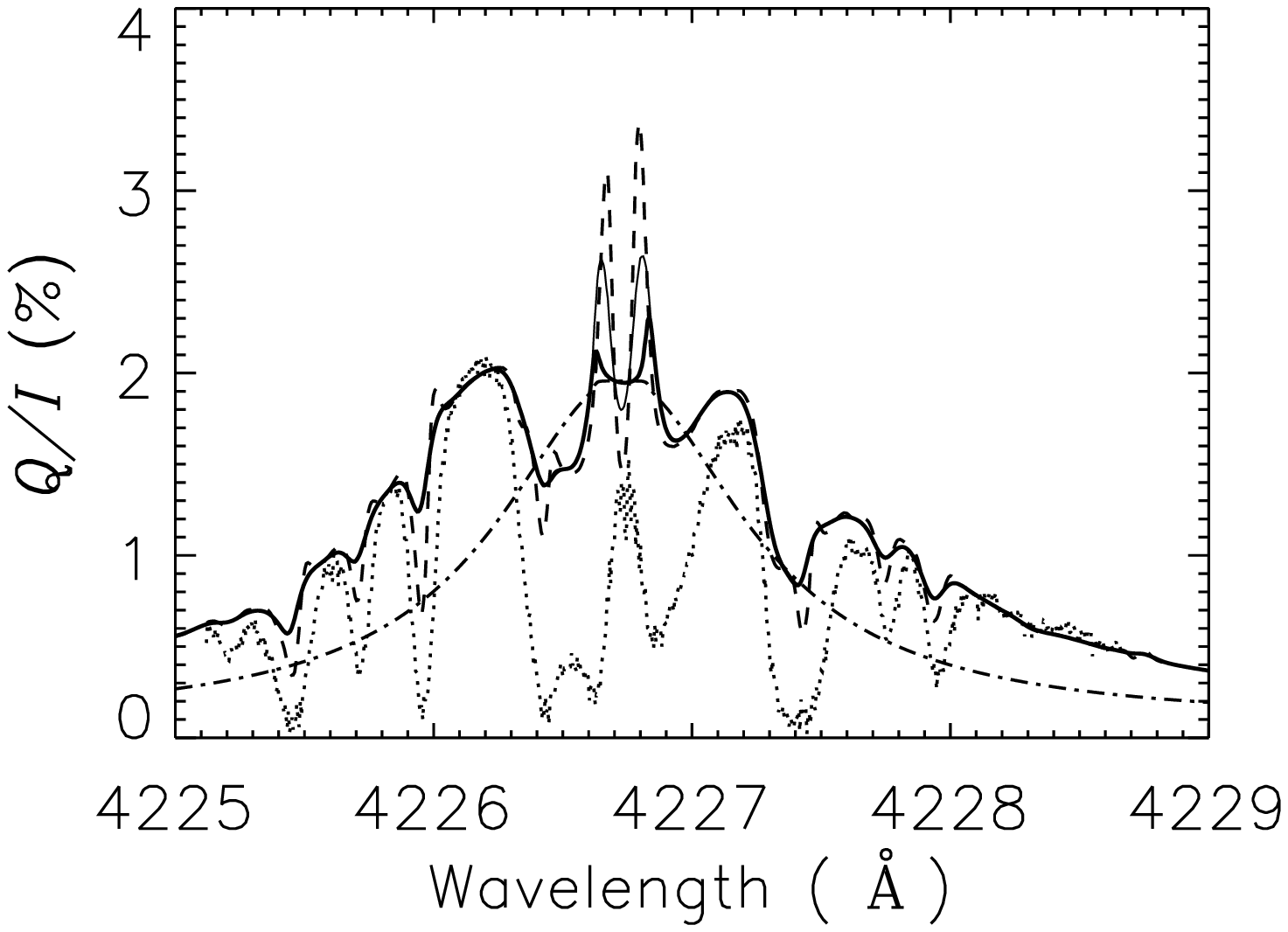}
\caption{Role of PRD. The heavy solid and dotted lines are the same 
as in Fig.~\ref{flatkgvskg}. The dashed line is the model profile computed with frequency-coherent scattering (CS), while the dot-dashed line is computed 
assuming CRD. The thin solid line is the model profile with $R_{\rm II-AA}$, while the heavy solid line has been computed with $R_{\rm II-AD}$. 
}
\label{prdonoff}
\end{figure}

In Fig.~\ref{prdonoff} we compare model profiles computed with different 
redistribution mechanisms. The coherent scattering (CS) limit can for 
example be obtained from the general PRD expression 
by choosing $R=R_{\rm CS}=H(\Delta\lambda,a)\,\delta(\lambda-\lambda^\prime)$ as the redistribution function. The angular dependence of the scattering process is then given by the non-magnetic 
Rayleigh phase matrix. One can see in Fig.~\ref{prdonoff} that except in the line core, where CS differs greatly from angle-dependent PRD 
(heavy solid line, which is the same as the solid line in 
Fig.~\ref{flatkgvskg}), pure coherent scattering (CS, dashed line) provides 
a good approximation at the Ca\,{\sc i} 4227\,\AA\ wing frequencies. In fact 
coherent scattering gives deeper minima in the blend lines than 
PRD. This is because when $C=0$ and $P_c=0$, the model profile obtained 
with coherent scattering exactly mimics the $k_{\rm G,\lambda}(\mu)$ spectrum, while PRD modifies it significantly in the inner core of the Ca\,{\sc i} 
4227 line and also in the cores of the blend lines (because although 
$R_{\rm II-AD}$ has coherent peaks in the wings, such peaks are not exactly 
delta functions unlike the case of pure coherent scattering and cause 
some broadening). 

Fig.~\ref{prdonoff} also shows a model profile computed with the assumption of CRD (dot-dashed line). We recover the CRD limit from the general expression by using $R=R_{\rm CRD} = H(\Delta\lambda^\prime,a)H(\Delta\lambda,a)$ for the redistribution function. The shape of the model profile obtained with CRD can be easily understood from Eqs.~(\ref{qbyi_nonmag_model}) and (\ref{pqline_prd}). In CRD the redistribution matrix elements are $R_{21} = 0.7425\, R_{\rm CRD}$, and $R_{11} = 1.2425\, R_{\rm CRD}$ \citep[see eg.][for the Rayleigh phase matrix expression]{nag03}. 
Thus $P_{Q,\rm line}$ becomes wavelength independent and is given by 
\begin{equation}
\label{pqline_crd}
P_{Q,\rm line} = 0.6\, {\int H(\Delta\lambda^\prime,a)\, 
k_{\rm G,\lambda^\prime}(\mu)\, I_{\lambda^\prime}(\mu=1)\,{\rm d}\lambda^\prime \over \int H(\Delta\lambda^\prime,a)\, I_{\lambda^\prime}(\mu=1)\,
{\rm d}\lambda^\prime}.
\end{equation}
Therefore the line contribution becomes constant at a value that is found to be 6\,\%. The shape of the CRD model profile is then entirely due to the factor 
$H(\Delta\lambda,a)/(H(\Delta\lambda,a)+C)$ in Eq.~(\ref{qbyi_nonmag_model}). 
As $C \ll 1$, this factor is close to unity at line center, where 
we have $(Q/I)_{\Delta\lambda=0} \approx 6\,\%\ \times\ S$. 
For the value of $S=0.325$ derived from the best fit (obtained with PRD), 
$(Q/I)_{\Delta\lambda=0} \approx 2\,\%$. As we move away from the line 
center, the factor $H(\Delta\lambda,a)/(H(\Delta\lambda,a)+C)$ decreases 
monotonically, and there is no possibility of modeling the observed $Q/I$ 
maxima of the Ca\,{\sc i} 4227 line anywhere in the line wings. 
We emphasize that it is not only the anisotropy 
of the incident radiation field that governs the shape and magnitude of 
the wing maxima in $Q/I$, but also a realistic redistribution mechanism, namely PRD. 

We also show in Fig.~\ref{prdonoff} a comparison between the model $Q/I$ profile based on the angle-averaged $R_{\rm II-AA}$ (thin solid line) and angle-dependent $R_{\rm II-AD}$ (heavy solid line) PRD mechanisms. Both succeed well in modeling the wing peaks within our LSA framework. The 
differences are noticeable only in the line core. The double peak in the line core has also been seen in the radiative transfer modeling by \citet{reneetal05}, who use $R_{\rm II-AA}$. Note that the same qualitative features of 
$(Q/I)_{\rm line-core}$ that we have found here for different line scattering mechanisms (CS, CRD, $R_{\rm II-AA}$, and $R_{\rm II-AD}$) may not be reproduced with the same details in full scale radiative transfer modeling. 

Comparison with the observed $Q/I$ spectrum shows that LSA allows us to model the $Q/I$ wings extremely well. With LSA we can fit the envelope (above the blend line depressions) of the observed $Q/I$. The line core of Ca\,{\sc i} 4227 
is however not modeled so well by LSA, although correct qualitative features like the $Q/I$ dips around the Doppler core are reproduced. This is also the case after including the contribution from $R_{\rm III-AD}$ type scattering through the introduction of elastic collisions (see below). This means that LSA does not work well enough in the line core, and that one may need radiative-transfer physics to explain the core shape of $Q/I$. This question is something we like to pursue in a future work by modeling the Ca\,{\sc i} 4227\,\AA\ line with full 
radiative transfer, to allow us to clarify and identify what aspect of radiative transfer is the source of the difference that we see in the line core.

The fit of the observations with the last scattering model leads to a scaling factor $S$ of 32.5\,\%\ rather than unity, which may be due to collisional depolarization, although radiative transfer effects may also cause deviations from LSA. If the main contribution to the scaling parameter $S$ comes from elastic collisions, then the scaling parameter gives us an estimate of the elastic collision rate $\Gamma_{\rm E}$. In the following section we discuss the effect of 
$\Gamma_{\rm E}$ on the model profiles, as it plays an important role in the modeling.  

\subsubsection{Role of elastic collisions $\Gamma_{\rm E}$}
In the previous sections we considered only limiting cases of frequency redistribution, namely frequency coherent scattering in the atomic frame (i.e., pure $R_{\rm II-AD}$ in the laboratory frame) and CRD. We now consider more realistic situations, where both types of scattering may occur, i.e., a weighted combination of $R_{\rm II-AD}$ and $R_{\rm III-AD}$ type scattering. The weights are the 
branching ratios given by
\begin{equation}
A = {\frac{\Gamma_{\rm R}}{\Gamma_{\rm R}+\Gamma_{\rm I}+\Gamma_{\rm E}}},
\label{brancha}
\end{equation}
for coherent scattering (in the atomic frame -- $R_{\rm II-AD}$ type), and 
\begin{equation} 
B^{(K)} = {\frac{\Gamma_{\rm E}-D^{(K)}}{\Gamma_{\rm R}+\Gamma_{\rm I}+\Gamma_{\rm E}}}\,
{\frac{\Gamma_{\rm R}}{\Gamma_{\rm R}+\Gamma_{\rm I}+D^{(K)}}},
\label{branchb}
\end{equation}
which represents the fraction of the scattering processes for which 
the atom is subject to elastic collisions that destroy the frequency 
coherence, but not the $2K$-multipole atomic polarization. 
Here $D^{(K)}$ is the rate of destruction of the 2$K$-multipole, with 
$K=0,1,2$ (note that $D^{(0)}=0$). As we are only considering 
linear polarization, only $D^{(2)}$ is relevant. $D^{(2)}$ is related 
to $\Gamma_{\rm E}$ through $D^{(2)}={\rm constant}\ \times\ \Gamma_{\rm E}$. The classical value of this constant is 
0.5 \citep[see][]{ste94}. Using an accurate form for the inter-atomic potential, \citet[][see also Faurobert-Scholl 1996]{fauetal95} estimate this constant to be 0.6 for the Ca\,{\sc i} 4227 line. In our modeling we therefore use $D^{(2)}=0.6\ \Gamma_{\rm E}$.

For the Ca\,{\sc i} 4227\,\AA\ line $\Gamma_{\rm E}$ is due to 
collisions with neutral hydrogen \citep[see][]{aueretal80,fau92}. The effect 
of $\Gamma_{\rm E}$ on $Q/I$ is to reduce the polarization at all wavelengths, 
and thereby make the scaling parameter $S$ become unity. Thus an increase in the elastic collision rate $\Gamma_{\rm E}$ causes a depolarization throughout the line profile. The effect of $D^{(2)}$ is limited to the line core, 
where it is somewhat similar to that of $\Gamma_{\rm E}$, but it 
does not affect the line wings, in contrast to $\Gamma_{\rm E}$. 
\begin{table}
\caption{The free parameters $S,\ C,\ P_c$ determined from the model 
fit for different choices of $\Gamma_{\rm E}/\Gamma_{\rm R}$. 
\label{free_para_model}}
\begin{center}
\begin{tabular}{crrrrrrrrrrr}
\tableline\tableline
$\Gamma_{\rm E}/\Gamma_{\rm R}$ & $S$ & $C$ & $P_c$ (\%) \\
\tableline 
         0       & 0.325 & 9.7E-5  & 0.25   \\
        0.5      & 0.345 & 1.5E-4  & 0.25   \\
         1       & 0.38  & 2.2E-4  & 0.25   \\
         2       & 0.43  & 3.5E-4  & 0.25   \\
         3       & 0.485 & 5.0E-4  & 0.25   \\
         5       & 0.67  & 1.0E-3  & 0.23   \\
        10       & 1.00  & 2.1E-3  & 0.13   \\
\tableline
\end{tabular}
\end{center}
\end{table}

\begin{figure}[ht]
\plotone{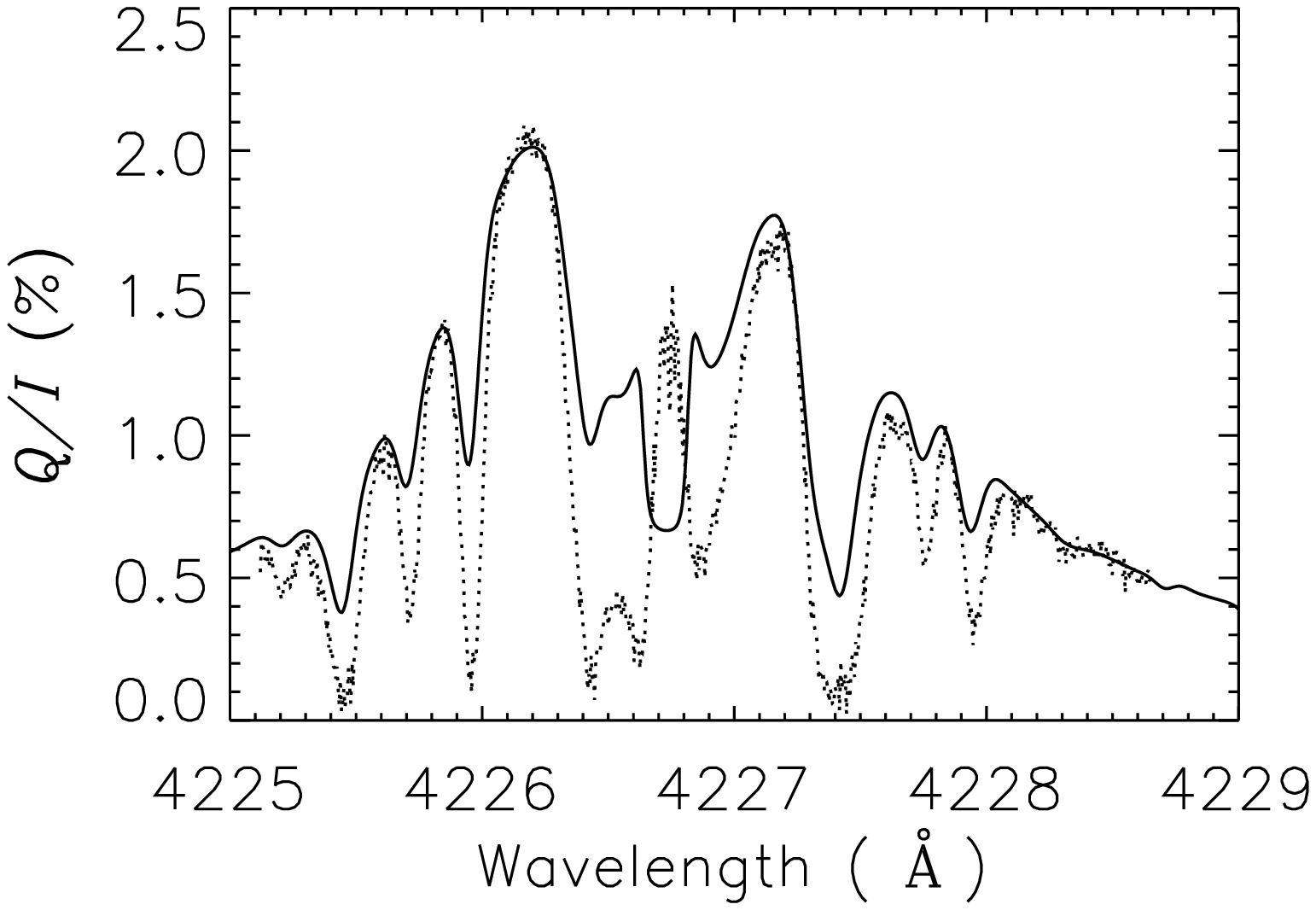}
\caption{Model fit obtained for $\Gamma_{\rm E}/\Gamma_{\rm R}=10$ (solid 
line). The dotted line is the observed $Q/I$. Note how well the observed 
$Q/I$ wings are fitted by the computed model profile. 
}
\label{model_fit_colli}
\end{figure}

With $\Gamma_{\rm E}$ as a free parameter, we have determined by model fitting the parameters $S,\ C,\ P_c$ for different choices of $\Gamma_{\rm E}$. The combination of free parameters thus determined are listed in Table~\ref{free_para_model}. We note that as $\Gamma_{\rm E}/\Gamma_{\rm R}$ increases, the continuum opacity parameter $C$ increases as well, and the scaling parameter $S$ approaches unity. The continuum polarization $P_c$ remains nearly constant. Furthermore, introduction of elastic collisions improves the fit to the observed $Q/I$ spectrum, in particular around the red wing maximum, and also the computed blend line minima become deeper as compared with the pure $R_{\rm II-AD}$ model fit (compare the solid lines in Figs.~\ref{flatkgvskg} and \ref{model_fit_colli}). For illustration we present in Fig.~\ref{model_fit_colli} an example of a model fit obtained for $\Gamma_{\rm E}/\Gamma_{\rm R}=10$. 

\subsection{An attempt to model the $(Q/I,U/I)$ wing signatures}
Next we try to model the $U/I$ observation shown in Fig.~\ref{obs_res2} with the modeling procedure described in \S~\ref{sec_hanle_model}. Observations show wing maxima in the $-U/I$ spectrum around 4226.2\,\AA\ and 4227.2\,\AA, which 
correspond approximately to $\pm 15$ Doppler widths from the 4226.74\,\AA\ line center. Note that the observed $U/I$ spectrum happens to be negative 
in our present recordings. In general $U/I$ spectra of either sign are 
equally likely, as they are due to a rotation angle that can be both 
positive and negative. In this paper we refer to the `$-U/I$ spectrum' 
to avoid confusion when we speak about polarization maxima in the line wings. 
Ambiguity might arise if we would instead speak of wing minima in $U/I$, since the absolute value of the polarization always has wing {\it maxima} in both the $+U/I$ and $-U/I$ cases. 

The parameters required for the modeling are Hanle $\Gamma_B = g e B/(2 m \Gamma_{\rm R})$ in standard notation, and $(\vartheta_B,\varphi_B)$ representing the orientation of a directed magnetic field, defined with respect to the vertical direction in the atmosphere. The elastic collision rate $\Gamma_{\rm E}$ is also used as a free parameter. We recall that the parameters $S$, $C$, and $P_c$ determined from the non-magnetic model fit for a given choice of $\Gamma_{\rm E}/\Gamma_{\rm R}$ (as listed in Table~\ref{free_para_model}) are kept constant when the magnetic field parameters are varied. In this way we have attempted to model the observed $(Q/I,U/I)$ spectra outside the so called ``non-magnetic'' 
regions, and at the spatial locations where the $(Q/I,U/I)$ wing signatures are seen. 

\subsubsection{Wing peaks of $Q/I$}
The framework that has been developed in the previous sections for modeling the non-magnetic $Q/I$ can still be used in the magnetized case with the modifications described in \S~\ref{sec_hanle_model}. For the purpose of discussion we introduce a quantity $\Delta(Q/I) = (Q/I)_{\rm mag} - (Q/I)_{\rm non-mag}$, which is a measure of the depolarization caused by the combined effect of magnetic and collisional depolarization that we get in PRD. In the top panel of Fig.~\ref{obs_res2} we observe depolarization in $Q/I$ (with respect to the non-magnetic $Q/I$), not only in the line core, but also in the wings (compare the solid and dashed lines in that figure). To model these observations (the dashed line in Fig.~\ref{obs_res2}), we varied the field parameters $(\Gamma_B,\vartheta_B,\varphi_B)$ and the elastic collision strength $\Gamma_{\rm E}/\Gamma_{\rm R}$. Our study shows that with the choice of pure $R_{\rm II-AD}$ to represent the PRD mechanism we do not get any wing depolarization ($\Delta(Q/I) \approx 0$), regardless of the choice of the field parameters. However, when we introduce elastic collisions we find that for an optimum choice of the combination $(\Gamma_B,\Gamma_{\rm E}/\Gamma_{\rm R})$, we do get wing depolarization ($\Delta(Q/I) \ne 0$) in $Q/I$. Let us next discuss briefly a few interesting aspects of this study. All the tests have been made for the field strength range $0.3 \le \Gamma_B \le 10$. 

\begin{figure}[ht]
\plotone{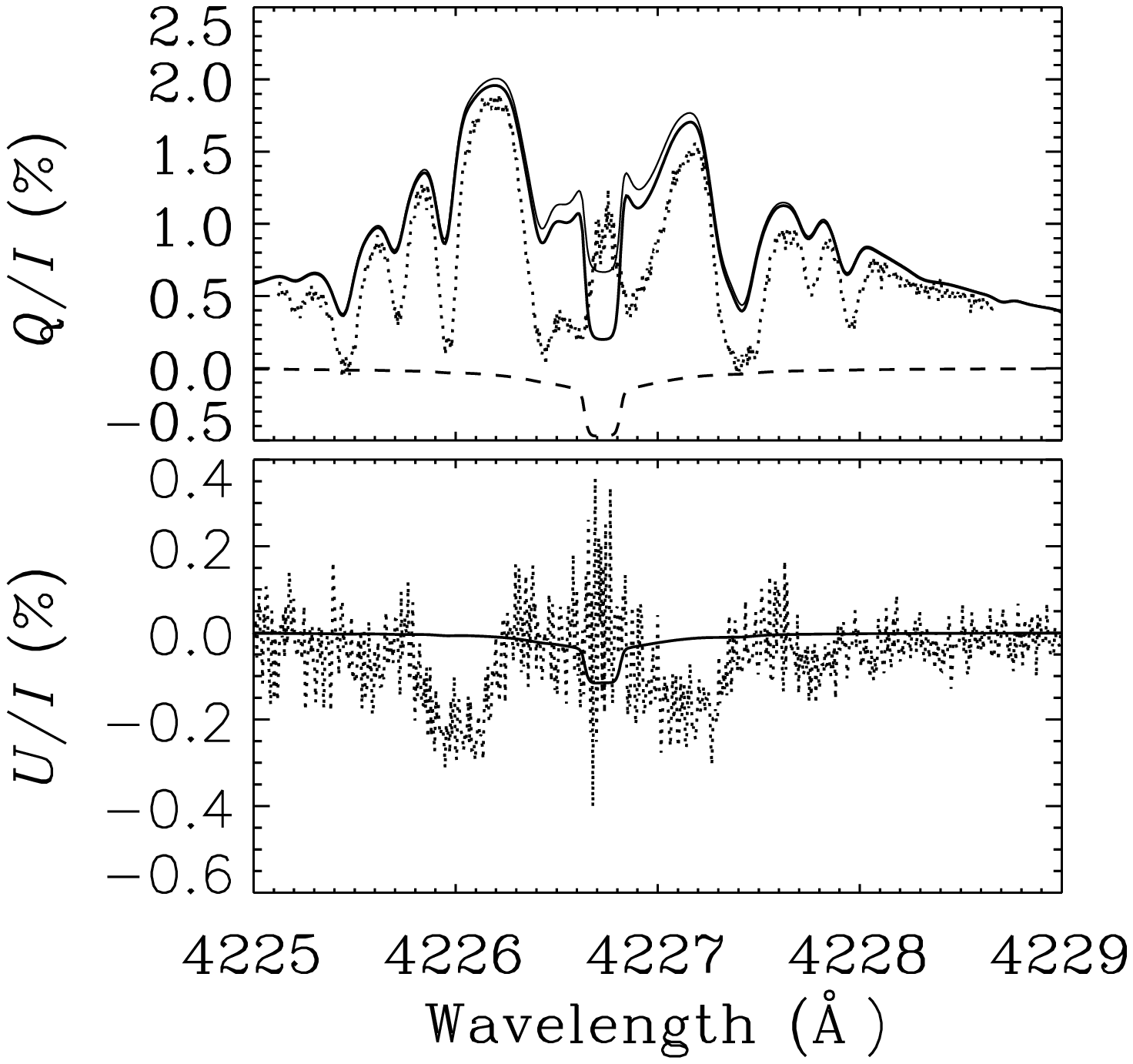}
\caption{Attempt to model the $(Q/I,U/I)$ wing signatures. The observations (dotted lines) shown here correspond to the ``magnetic observations'' presented in Fig.~\ref{obs_res2}. In the $Q/I$ panel the heavy solid line represents the magnetic model profile, the thin solid line the non-magnetic model profile, and the dashed line their difference. In the $U/I$ panel the solid line represents the magnetic model profile. The parameters used are  $(\Gamma_B,\vartheta_B,\varphi_B; \Gamma_{\rm E}/\Gamma_{\rm R}) =(10, 90^\circ, 135^\circ; 10)$.  
}
\label{model_ubyi}
\end{figure}

In the line core $\Delta(Q/I)$ decreases towards zero as 
$\Gamma_{\rm E}/\Gamma_{\rm R}$ increases. This shows that for large values of 
$\Gamma_{\rm E}/\Gamma_{\rm R}$ the effect of collisional depolarization 
dominates over the effect of magnetic depolarization in the line core. In the 
line wings $\Delta(Q/I)\approx 0$ when $\Gamma_B < 3$. For $3 \le \Gamma_B \le 10$ the wing signature $\Delta(Q/I)$ initially increases slowly with $\Gamma_{\rm E}/\Gamma_{\rm R}$, and then decreases towards zero with a further increase of the collision strength. For example, when $(\Gamma_B,\Gamma_{\rm E}/\Gamma_{\rm R})=(10,10)$ we observe Hanle depolarization that extends into the wings as shown in Fig.~\ref{model_ubyi} (compare the heavy solid and the thin solid lines), but no wing peaks are obtained in $\Delta(Q/I)$ that is represented by the dashed line. If we further increase $\Gamma_{\rm E}/\Gamma_{\rm R}$ (say to 100), then $\Delta(Q/I)$ vanishes in the line wings and becomes almost zero also in the line core, implying that the collisional depolarization effect again dominates over the magnetic depolarization effect. These two competing effects together decide the extent of depolarization in the wings. 

\subsubsection{Wing peaks of $U/I$}
Our modeling efforts turn out to be unsuccessful in reproducing the observed wing maxima in $-U/I$, contrary to our expectations. We expected that the elastic collisions play a significant role in transferring the Hanle effect from the line core to the line wings without destroying the atomic polarization \citep{nagetal03,bia03,sametal07b}. This expectation is satisfied to some extent for $Q/I$ for an optimum choice of the parameter pair $(\Gamma_B,\Gamma_{\rm E}/\Gamma_{\rm R})$, but even in this optimized case we fail to reproduce the wing 
maxima that are observed in $\Delta(Q/I)$. In the case of $U/I$ we find that for $\Gamma_B < 3$ the amplitude of $U/I$ in the line core gradually decreases when 
$\Gamma_{\rm E}/\Gamma_{\rm R}$ increases, but correspondingly no wing peaks appear at 
all. For $ 3 \le \Gamma_B \le 10$, $U/I$ in the line core initially increases slightly with $\Gamma_{\rm E}/\Gamma_{\rm R}$ but then decreases with the further increase in the collision strength. Again no wing peaks appear. Variations of the magnetic field parameters also do not help to reproduce the $U/I$ wing peaks (like they failed to reproduce the $\Delta(Q/I)$ wing peaks). The line core peak in $U/I$ on the other hand sensitively responds to variations of all the above free parameters, in a manner that is well understood \citep{nagetal02}. 

\section{Conclusions}
\label{sec_conclu}
In the present paper we have developed a simple framework based on the 
last scattering approximation (LSA) to model the Second Solar Spectrum. This approximation gives excellent fits to the linear polarization that is observed in the wings of spectral lines, as demonstrated for the case of the Ca\,{\sc i} 4227\,\AA\ line. However, fitting the line core polarization may require 
the solution of the polarized radiative transfer equations (including PRD), at 
least for strong resonance lines. The most important quantity in our modeling 
is the anisotropy factor $k_{\rm G,\lambda}(\mu)$, which we determine from the 
observed center-to-limb variation of the Stokes $I$ spectrum. The detailed 
wavelength variation of the limb-darkening function plays a fundamental role and is responsible in particular for the occurrence of $Q/I$ minima that surround the core region and separates it from the wing maxima. Another key ingredient is the appropriate partial frequency redistribution matrix to be used. The detailed validity range of the last scattering approximation (extent of its applicability in the line cores of strong lines and throughout the line profiles in the case of weak lines) needs to be explored by benchmark tests with full scale radiative transfer, before the diagnostic potential of this approach can be fully exploited. 

We have applied the LSA framework to explore the question whether the observed spatial variations in the $Q/I$ and $U/I$ wings of the Ca\,{\sc i} 4227\,\AA\ line may be explained in terms of the Hanle effect, which usually is confined to the Doppler core of spectral lines but could in principle become active in the far line wings through frequency redistribution mediated by elastic collisions. Such Hanle-like wing signatures were noticed for the first time in active regions by \citet{bia03}, but in the present paper we report observations showing that these wing signatures are present in quiet solar regions as well. Our attempts to model these $(Q/I,U/I)$ wing signatures failed to reproduce them. Both the 
$\Delta(Q/I)$ profile of the $Q/I$ spatial variations and the $-U/I$ profile 
are observed to have maxima in the wings, similar in shape to the $Q/I$ 
non-magnetic profile. However the $\Delta(Q/I)$ and $-U/I$ modeling 
failed to retrieve this property, although we searched the whole parameter 
space of collision rates and magnetic-field parameters. 

This null result appears to rule out a direct magnetic-field origin (via the Hanle effect) of the observed spatial variations of the scattering polarization in the line wings, in contradiction to earlier suggestions 
\citep{nagetal03,bia03,sametal07b}, at least within the framework of the currently 
available PRD theory. This points in the direction of a non-magnetic interpretation, 
which may include local deviations from a plane-parallel stratification 
\citep[see][for some information on the possible effects]{rj99}
with an inhomogeneous solar atmosphere containing ``hot spots'' \citep[see][who used this mechanism to interpret the $Q/I$ and $U/I$ fluctuations seen abundantly in the Ca K line]{renesten07}. The local density inhomogeneities may also cause significant fluctuations of the collisional depolarization rate. Detailed analysis of these possibilities is outside the scope of the present paper, but such alternative interpretations clearly need to be explored. Ideally one would like to do 2-D mapping of the Stokes vector (rather than work with single slit positions) with high spatial resolution to map the polarization signatures together with the intensity structures to examine whether the non-magnetic interpretation is viable. 

The non-magnetic interpretation is not without its own problems. Thus 
with a simple-minded model for spatially varying deviations from a 
plane-parallel stratification one would expect a spatial correlation 
between the line wing fluctuations seen in $Q/I$ and $U/I$, but such a 
correlation seems to be weak or nearly absent in our observations. This 
indicates that the 3-D atmospheric structuring that one would need is 
more complex, and that we may have to consider a 
mixture of geometry and collisional effects. It is important to quantify 
how much the atmosphere deviates from a plane-parallel stratification in 
lines of various strengths, and how these deviations are spatially 
structured, and coupled to the opacity structure within the line as we 
move from the core to the wings of the line. While this is a challenging 
problem, it is within reach with the new generation of observing facilities 
that are becoming available. The problem can also be approached by 
numerical simulations to generate 3-D atmospheric models, and then use 
3-D radiative transfer to compute the linearly polarized line profiles 
that emerge from this atmosphere \citep[e.g., as done by][for the Sr\,{\sc i} 
4607\,\AA\ line assuming CRD]{jtbetal04,js07}. Such a project for the 
Ca\,{\sc i} 4227\,\AA\ line for which PRD effects are important would be extremely 
demanding on computing resources, but it is something that also will 
soon be within reach. 

\acknowledgments
K.N.N. and M.S. would like to thank IRSOL for supporting their visits 
during 2004, 2005 and 2007 in connection with this project. 
IRSOL is financed by Canton Ticino, ETHZ and City of Locarno together with
the municipalities affiliated to CISL. Scientific projects are supported
with SNF grant 200020-117821.


 
\end{document}